\documentclass{llncs} %

\usepackage{float}
\usepackage{graphicx}
\usepackage{wrapfig}

\usepackage[labelfont=bf]{caption}
\usepackage{amsmath} 
\usepackage{mathtools} 
\usepackage{amsfonts} 
\usepackage{amssymb}

\usepackage{array}
\usepackage{booktabs,arydshln}

\usepackage{rotating}
\usepackage{multirow}
\usepackage{multicol}
\usepackage{lscape}
\usepackage{subfig}

\usepackage[export]{adjustbox}

\usepackage{algorithm,algorithmicx,algpseudocode}
\usepackage[colorlinks=true,linkcolor=blue]{hyperref}%

\usepackage{color,soul}

\usepackage{color}

\definecolor{darkgreen}{rgb}{0.53, 0.66, 0.42}

\begin{document}

\title{A Review on Image- and Network-based Brain Data Analysis Techniques for Alzheimer's Disease Diagnosis Reveals a Gap in Developing Predictive Methods for Prognosis}
\titlerunning{Short Title}  

\author{Mayssa Soussia\index{Soussia, Mayssa} \and Islem Rekik\index{Rekik, Islem}\thanks{Corresponding author: {irekik@dundee.ac.uk}, \url{www.basira-lab.com}} }

\authorrunning{M. Soussia et al.}   

\institute{BASIRA lab, CVIP group, School of Science and Engineering, Computing, University of Dundee, UK \ }

\maketitle              

\begin{abstract}
	
Unveiling pathological brain changes associated with Alzheimer's disease (AD) is a challenging task especially that people do not show symptoms of dementia until it is late. Over the past years, neuroimaging techniques paved the way for computer-based diagnosis and prognosis to facilitate the automation of medical decision support and help clinicians identify cognitively intact subjects that are at high-risk of developing AD. As a progressive neurodegenerative disorder, researchers investigated how AD affects the brain using different approaches: 1) \emph{image-based methods} where mainly neuroimaging modalities are used to provide early AD biomarkers, and 2) \emph{network-based methods} which focus on functional and structural brain connectivities to give insights into how AD alters brain wiring. In this study, we reviewed neuroimaging-based technical methods developed for AD and mild-cognitive impairment (MCI) classification and prediction tasks, selected by screening all MICCAI proceedings published between 2010 and 2016. We included papers that fit into image-based or network-based categories. The majority of papers focused on classifying MCI vs. AD brain states, which has enabled the discovery of discriminative or altered brain regions and connections. However, very few works aimed to \emph{predict} MCI progression based on early neuroimaging-based observations. Despite the high importance of reliably identifying which early MCI patient will convert to AD, remain stable or reverse to normal over months/years, predictive models are still lagging behind.

\end{abstract}

\section{Introduction}

Alzheimer's disease (AD), which is the most common form of dementia, is still today an incurable degenerative disease. It affects 5-8\% of all people above 60 years of age, increasing to around 40\% of people older than 90\% \cite{report2015}. AD is also known as an irreversible, progressive disorder that destroys neurons which leads to deficits in cognitive functions such as memory and thinking skills. Clinical diagnosis can be supported by biomarkers that detect the presence or absence of the disease. 
However, identifying such biomarkers, especially in a very early stage,  remains challenging as brain changes due to AD occur even before amnestic symptoms appear \cite{buckner2004}.
The number of people diagnosed with dementia in the UK is expected to rise to over 2 million by 2051 with an estimated cost at between £17 billion and £18 billion a year (Dementia UK report\footnote{\url{https://www.alzheimers.org.uk/about-us/policy-and-influencing/dementia-uk-report}}). Hence, identifying Alzheimer's disease (AD) earlier before the neurodegeneration is too severe and where treatment is not currently available, might aid in preventing AD onset. Specifically, patients initially diagnosed with mild cognitive impairment (MCI) are known to be a clinically heterogeneous group with different patterns of brain atrophy \cite{Misra:2009}, of which some cases will not progress to AD  \cite{Bron:2015}. To examine the borders between MCI and AD, Magnetic Resonance Imaging (MRI) was extensively used as a non-invasive imaging modality to track changes in brain images of MCI patients as they remain stable, progress to AD, or reverse to normal. Brain dementia MRI data are rapidly growing with emerging international research initiatives aiming to massively collect large high-quality brain images with structural, diffusion and functional imaging modalities, e.g., the public ADNI (Alzheimer's Disease Neuroimaging Initiative) dataset \cite{Jack:2008}. However, despite the large body of publications on AD and major advances in neuroimaging technologies, brain image analysis and machine-learning methods, dementia research has not progressed as desired. Fundamentally, there are two major reasons for this.
					
\emph{First}, the majority of methods developed for investigating AD stages have focused on learning how to classify AD vs. MCI or normal control (NC) subjects 
\cite{Coupe:2011,Liu:2013b,Suk:2013,Jie:2013,Suk:2014,Min:2014,An:2016,Peng:2016,Liu:2013a,Liu:2016a,Liu:2016b,Wee:2011,Wee:2012,Wee:2013,Jie:2014,Suk:2015,Chen:2016,Yu:2016,Leung:2010}. A conventional classification method would help identify features discriminating between MCI and AD groups; however, it would not allow to identify MCI patients with longer-term followup who will convert to AD after the first MR acquisition timepoint (i.e., baseline). Recently, a challenge on computer-aided diagnosis (CAD) of dementia based on structural MRI, namely CAD-Dementia \cite{Bron:2015}, was launched to evaluate the performance of 29 algorithms from 15 research teams in classifying NC/MCI/AD using a public dataset. However, such dementia challenges have not focused on finding very early biomarkers of prodromal AD, characteristic of the presymptomatic MCI phase of the disease preceding severe cognitive decline, which is a major issue for current international research on AD.

\emph{Second}, although advanced machine-learning and medical imaging analysis methods for dementia CAD have demonstrated high performance in the literature \cite{Bron:2015}, they are not publicly shared for comparability, reproducibility, and generalizability to unseen data \cite{nichols2017}. Although the data for the CAD-dementia challenge is available, the developed methods were not made available for researchers to test on other datasets. A notable exception based on multivariate analysis \cite{Sabuncu:2015} overlooks the richness and efficiency of recently published machine-learning and data analysis methods for brain disease diagnosis and prognosis \cite{Brown:2016}.
In the following sections, we provide in-depth analysis of AD-related classification and evolution prediction methods from various neuroimaging modalities and identify the gaps in the state-of-the-art.

\section{Selection criteria}
\begin{wrapfigure}{l}{0.5\textwidth} 
\centering
\includegraphics[width=0.4\textwidth]{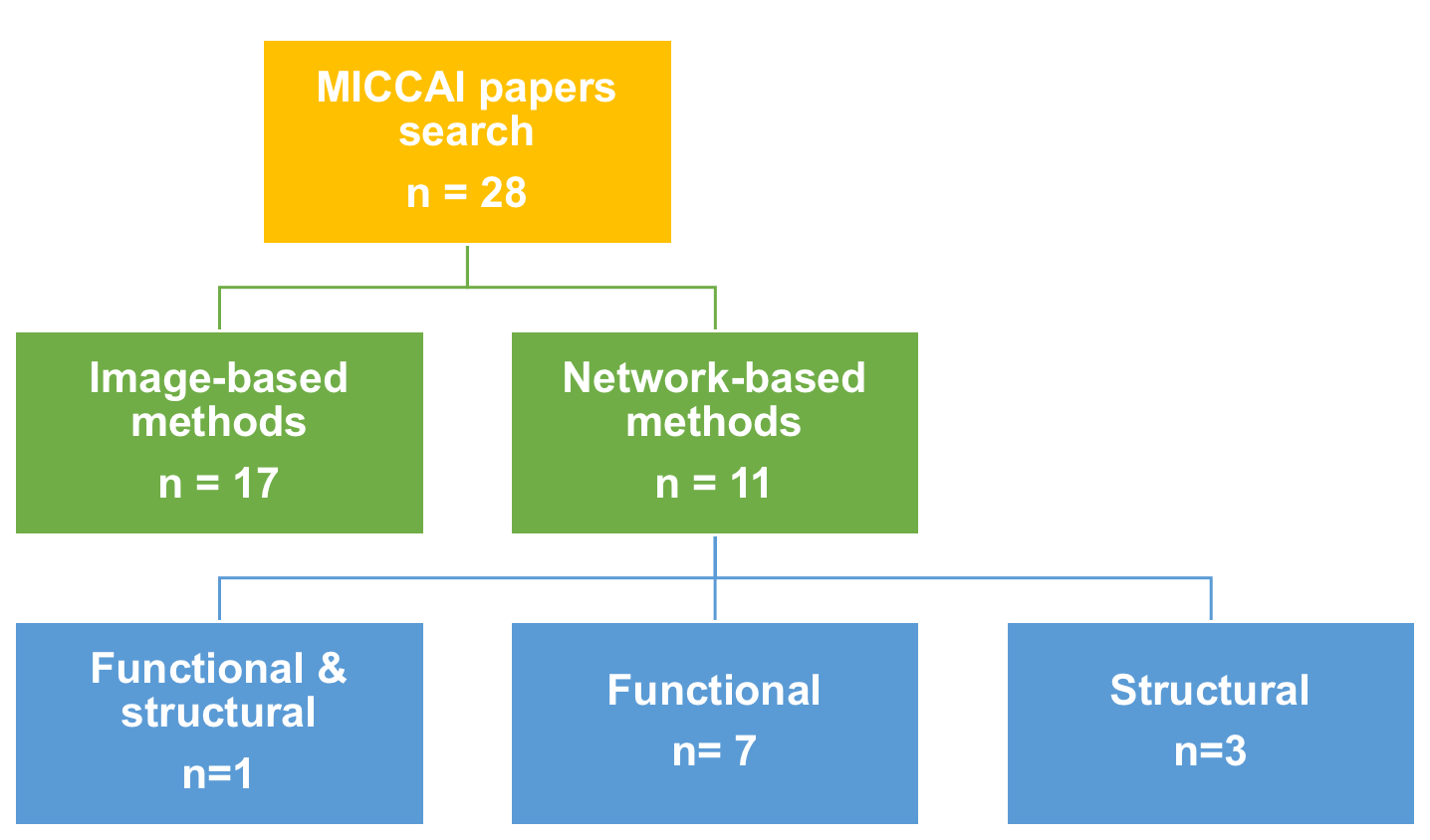}
\caption{\emph{Identified AD/MCI classification papers published in MICCAI 2010--2016 proceedings.}}
\label{fig:0}
\end{wrapfigure}

The analyzed papers in this review were identified from MICCAI conference from 2010 until 2016. We conducted our search using different combinations of the following key words: \emph{functional, structural, fMRI, DTI, magnetic resonance imaging, network, brain, connectivity, diffusion, Alzheimer's disease, mild cognitive impairment, classification, prediction, diagnosis, AD, MCI, biomarker, dementia}. We identified 28 papers based on the given search criteria. It is noteworthy that works developed for segmentation tasks and those not focusing primarily on AD/MCI classification or prediction were excluded from this review. The included papers are displayed in \textbf{Fig.}~\ref{fig:0}. We grouped them into two categories: `image-based methods' and `network-based methods'.

\section{Image-based methods} 

We identified 17 papers that use MR images for dementia state classification. These mainly used hippocampal atrophy and gray matter volume for classifying NC, MCI, and AD brain states. This can be explained by the fact that AD is related to gray matter loss \cite{karas2003} and the shape of subcortical structures (particularly the hippocampus) \cite{du2001}. To predict clinical decline at the MCI stage and progression to AD, \cite{Leung:2010} created p-maps from the differences in the shape of the hippocampus between NC and AD subjects and showed increased rates by identifying local regions of interest (ROIs) within the hippocampus using statistical shape models. In a different work, \cite{Iglesias:2011} used (left and right) caudate nucleus, and putamen as additional features to hippocampal features to present a system for AD classification using a self-smoothing operator. Other papers suggested the combination of grading measure with HC volume. Specifically, \cite{Coupe:2011} proposed a new method to robustly detect the hippocampal atrophy patterns based on a nonlocal means estimation framework. Combined with HC volume, the grading measure (i.e, the atrophy degree in AD context) led to a success classification rate of 90\% between NC and AD subjects. 

\cite{Liu:2013b} used two different modalities (MRI + PET) where each subject is represented by two 93-dimensional feature vectors that represent gray matter volume and the average intensity of PET images of 93 ROIs. A novel multi-task learning based feature selection method was proposed to preserve the complementary information conveyed by the two modalities and reached 94\% accuracy in distinguishing between AD and NC subjects. In the same context, \cite{Jie:2013} proposed a manifold regularized multi-task learning framework to jointly select features from multi-modality (MRI + FDG-PET) data as well as \cite{Suk:2014} and \cite{zhu2014,An:2016,Peng:2016}.
Similarly, \cite{Suk:2013} used the same features in addition to three CSF biomarkers and introduced a deep learning method that discovers the non-linear correlations among features which improves the AD, MCI and MCI-C diagnosis accuracy. 		 	 	 		
In \cite{Liu:2013c}, a novel  Multifold Bayesian Kernelization (MBK) method was proposed to analyze multi-modal MRI biomarkers including average cerebral metabolic rate from PET data, gray matter volume, solidity and convexity features for AD and MCI classification.
Another study \cite{Min:2014} introduced a different approach to improve AD/NC and p-MCI/s-MCI classification. Basically, it learns a maximum margin representation using multiple atlases jointly with the classification model which resulted in 90\% accuracy for AD/NC classification and 73\% for p-MCI/s-MCI.

\begin{wrapfigure}{r}{0.5\textwidth} 
\centering
\includegraphics[width=0.4\textwidth]{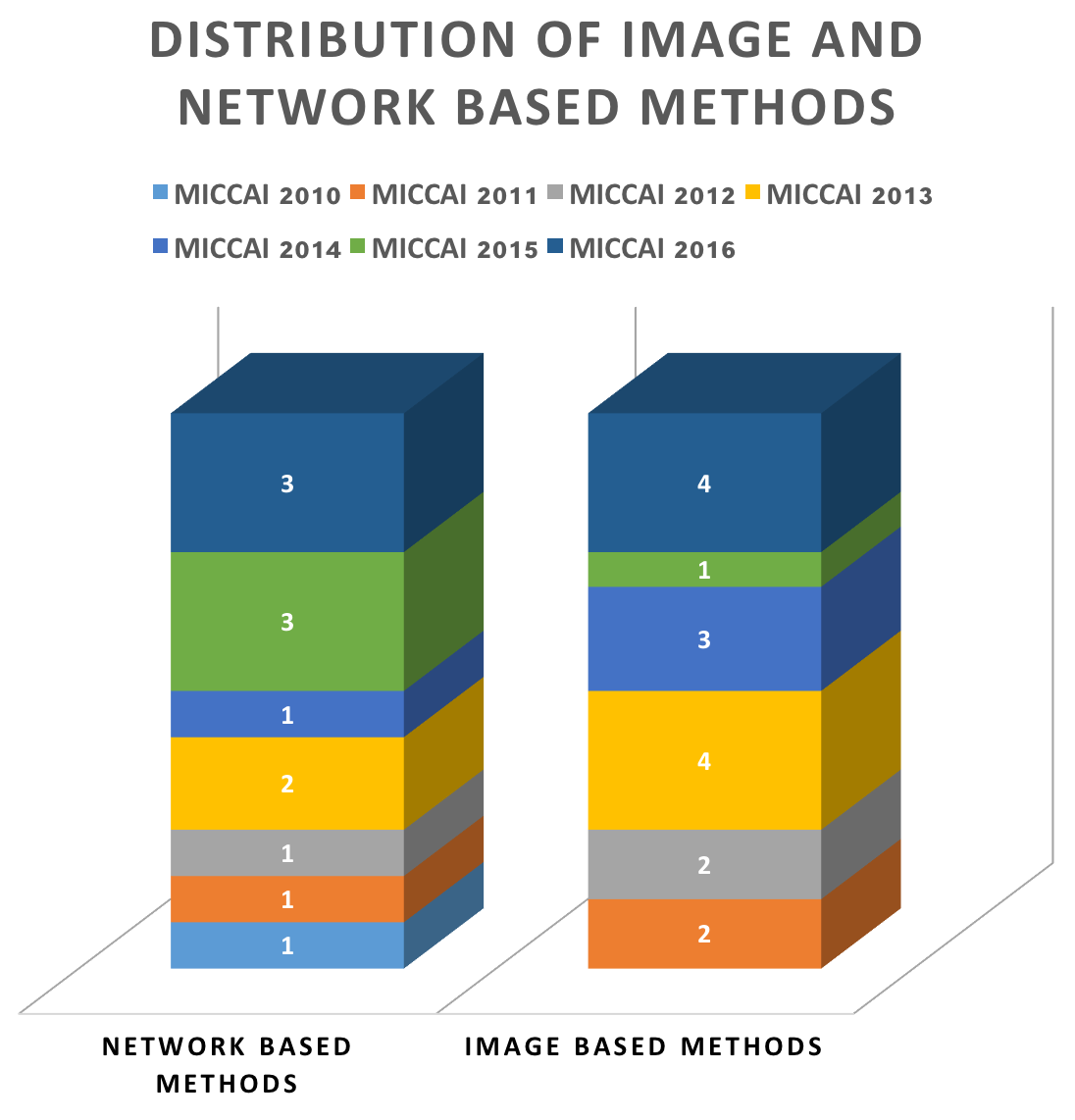}
\caption{\emph{Distributions of the identified image-based and network dementia diagnosis and prognosis methods published in MICCAI 2010-2016 proceedings.}}
\label{fig:1}
\end{wrapfigure}

We also identified two landmark papers \cite{Thung:2015,Zhu:2016}, which devised machine learning frameworks for \emph{predicting} dementia evolution at later timepoints. \cite{Zhu:2016} proposed a novel canonical feature selection method to fuse information from different imaging modalities (MRI+PET). Specifically, original features (gray matter volume and average intensity of PET images) are projected into a common space. Hence, they become more comparable and easier to depict their relationship in order to predict clinical scores of Alzheimer's disease. Using the same features, \cite{Thung:2015} applied a low rank subspace clustering to cluster the data, then used a low rank matrix completion framework to identify pMCI patients and their time of conversion.

\section{Network based methods}

In this section, we identified 12 papers where the majority (6 papers) used functional networks (derived from fMRI) for dementia diagnosis and prognosis. The remaining studies mainly used structural brain networks (derived from MRI) or fused both networks to enhance AD classification accuracy.

\subsection{Functional network}
It has been reported that the neurodegenerative process of AD reflects disturbed functional connectivity between brain regions \cite{fransson2005,wang2007}. These alterations are usually measured using resting-state functional MRI (rs-fMRI). Multiple methods have been used for AD diagnosis. For instance, \cite{Wee:2012} proposed a constrained sparse linear regression model associated with the least absolute shrinkage and selection operator (LASSO) that generates topologically consistent functional connectivity networks from rs-fMRI, thereby improving NC/MCI classification compared with traditional correlation-based methods \cite{wang2007,wee2012}. \cite{Wee:2013} introduced a sparse  multivariate autoregressive (MAR) modeling to infer effective connectivity from rs-fMRI data and demonstrated its superiority compared to correlation based functional connectivity approaches.

\cite{Yu:2016} proposed a novel weighted sparse group representation method for brain network modeling, which integrates link strength, group structure as well as sparsity in a unified framework. This models the interactions among multiple brain regions unlike pairwise Pearson correlation and showed superior results in the task of MCI and NC classification. Since  simply relying on pairwise functional connectivity between brain regions overlooks how their relationship might be affected at a higher-order level by AD, many studies introduced a functional connectivity hyper-network (FCHN) to infer additional information for AD classifcation.  \cite{Jie:2014} constructed a FCHN using sparse representation modeling where three sets of brain regions are exracted to be fed into a multi-kernel SVM classifier and evaluated on a real MCI dataset. \cite{Chen:2016b} also used a high-order functional connectivity network (HOFC) for MCI classification but generated multiple HOFC networks using hierarchical clustering to further ensemble them with a selective feature fusion method. This approach produced better classification performance than simple use of a single HOFC network.

\textbf{Dynamic FC.} While many studies assumed stationarity on the functional networks over time \cite{li2012,wee2014g}, recent studies in neuroscience have shown that functional organization changes spontaneously over time \cite{hutchison2013}. For instance, \cite{Suk:2015} introduced a novel method to model functional dynamics in rs-fMRI for MCI identification. Specifically, a deep network was designed to unravel the non-linear relationships among ROIs in a hierarchical manner and achieved a high classification performance.

\subsection{Structural network} 
Recent findings showed that AD induces a disrupted topology in the structural network characterized by an early damage to synapses and a degeneration of axons \cite{serrano2011}. These alterations are basically investigated using MRI with different proposed techniques. In \cite{Liu:2013a}, a graph matching framework was devised to match (i) a source graph, where each node encodes a vector that describes regional gray matter volume or cortical thickness features, and (ii) a target graph that includes class label and clinical scores. This approach estimates a target vector for each sample without neglecting its relation with other samples. \cite{Liu:2016a} proposed a view-aligned hypergraph learning (VAHL) method using multi-modality data (MRI, PET, and CSF)  for AD/MCI diagnosis where each view corresponds to a specific modality or a combination of several ones. This method can explicitly model the coherence among the views which led to a boost of 4.6\% in classification accuracy.
\cite{Gao:2015a} proposed a two-stage (query prediction + ranking) medical image retrieval technique with application to MCI diagnosis assistance. This framework was evaluated using three imaging modalities: T1-weighted imaging (T1-w), Diffusion Tensor Imaging (DTI) and Arterial Spin Labeling (ASL).

\section{Functional and structural networks} 
Relying on either structural or functional brain networks may overlook the complementary information that can be leveraged to improve diagnosis and prognosis. For this purpose, \cite{Wee:2011} integrated two imaging modalities (DTI and fMRI) using a multi-kernel support vector machine (SVM) to improve classification performance where DTI images are parcellated into 90 regions. Then, different structural networks are generated, each conveying a different biophysical property of the brain (e.g., fibers count, fractional anisotropy and mean diffusivity). Additionally, functional connectivity matrices were constructed based on Pearson correlation coefficient to encode the connectivity strength between a pair of ROIs. 
Furthermore, \cite{Gao:2015b}  proposed a centralized hypergraph learning method to model the relationship among subjects using multiple MRIs. Specifically, four MRI sequences were used including T1-weighted MRI (T1), Diffusion Tensor Imaging (DTI), Resting-State functional MRI (RS-fMRI) and Arterial Spin Labeling (ASL) perfusion imaging. This allows to extract supplementary information captured by different neuroimaging data, thereby enhancing the quality of MCI diagnosis.  	

\section{Results and discussion} 

In this paper, we identified and reviewed 28 works on AD diagnosis and prognosis published in MICCAI proceedings from 2010-2016. \textbf{Table.}~\ref{tab:0} displays the different identified papers, while revealing five major gaps that need to be addressed to move dementia research field forward. \emph{First,} all identified MICCAI papers focused on AD/MCI/NC classification, except for two papers \cite{Thung:2015,Zhu:2016}, which proposed machine-learning frameworks to predict MCI conversion to AD. Undeniably, accurate discrimination between AD and MCI subjects is an important task to solve as it helps devise more individualized and patient-tailored treatment strategies \cite{ithapu2015}. However, an accurate prognosis for MCI patients is far more important for providing the optimal treatment and management of the disorder in very early stage. Indeed, early biomarkers identification might help reduce MCI to AD conversion rate. Therefore, predictive models need to be developed to fill this gap and propel the field of MCI prognosis forward. 

Such lack of studies could be due to the scarcity of spatiotemporal neuroimaging data where each patient is scanned multiple times. One way to tackle this is by adopting good practices in data analysis and sharing which can promote reliability and collaboration \cite{nichols2017}. \emph{Second,} the classification performance of the proposed technical methods for dementia largely varied with multiple peaks and drops from 2010-2016. This can give insights into the heterogeneity and variability of the disease within subjects and how challenging it is to find an accurate method that works for all cases. In fact, no single approach can be sufficient as each has complementary merits and limitations. \emph{Third,} comparing these methods is very difficult since they used different approaches and datasets, it is somewhat hard to tell which one performs better if they are not compared against the same baseline methods and evaluated on the same dataset. \emph{Fourth,} all network-based analysis methods overlooked how dementia states affect the relationship between cortical regions \emph{in morphology} in both stability, conversion, or reversal MCI evolution scenarios. To fill this gap and noting that several studies \cite{Brown:2016,Querbes:2009} reported that morphological features of the brain, such as cortical thickness, can be affected in neurological disorders, one can use the recently proposed morphological brain networks for dementia diagnosis  \cite{Lisowska:2017,Mahjoub:2018,Lisowska:2018,Raeper:2018}. \emph{Last,} none of these works proposed a technique for predicting the full trajectory of brain shape changes as MCI progresses towards AD, remains stable, or reverses to normal. Besides, the absence of network-based predictive models is remarkable (\textbf{Table}~\ref{tab:0}). As such, the use of advanced network and shape analysis methods, using machine learning, could prove fruitful for both classification \cite{Lisowska:2017,Mahjoub:2018,Lisowska:2018,Raeper:2018} and prediction tasks.

\begin{sidewaystable}[thp]
\centering
\scalebox{0.54}{

\begin{tabular}{c@{~~}c@{~~}c}
	\toprule
	Data & Image-based MICCAI papers & Network-based MICCAI papers \\
	\midrule
	
	AD/NC & \cite{Iglesias:2011,Coupe:2011,Liu:2013b,Suk:2013,Jie:2013,Suk:2014,Min:2014,An:2016,Peng:2016} & 	\cite{Liu:2013a,Liu:2016a,Liu:2016b} \\
	
 		  & 96.25\%|90\%|94.37\%|95.9\%|95.03\%|95.18\%|90.69\%| 92.1\% |96.1\% & 92.17\%|93.10\%|94.05\% \\
    
   NC/MCI & \cite{Iglesias:2011,Liu:2013b,Suk:2013,Jie:2013,Suk:2014,An:2016,Peng:2016} & \cite{Wee:2011,Wee:2012,Liu:2013a,Wee:2013,Jie:2014,Suk:2015,Chen:2016,Liu:2016a,Yu:2016,Liu:2016b} \\

		  & 91.25\%|78.8\%|85\%|79.27\%|79.52\%|79.9\% |80.3\%| & 96.59\%|86.49\%|81.57\%|91.89\%|94.6\%|81.08\%|84.85 \%|80.00\%|81.8 \%|88.59\% \\
		
	cMCI/sMCI & \cite{Leung:2010,Cheng:2012,Singh:2012,Suk:2013,Jie:2013,Suk:2014,Min:2014,An:2016,Zhang:2016} & \cite{Liu:2016a} \\
			  & AUC(0.67) |69.4\%|66\%|75.8\%|68.94\%|72.02\%|73.69\%|80.7\%|96.7\%| & 79\% \\

	\cmidrule(lr){2-3}
	Prediction & \cite{Liu:2013c} & \\
	  		   &  NC: 86\% |  cMCI: 60.61\% | sMCI: 66.96\% | AD:81.76\% & \\
	
	\textcolor{red}{MCI conversion prediction using baseline MRI}  & \textcolor{red}{\cite{Thung:2015}} & \\
						 & \textcolor{red}{pMCI to AD: 76\% } 			 & \\
	
	\textcolor{red}{MCI conversion prediction using MRI}  & \textcolor{red}{\cite{Zhu:2016}} & \\
											 & \textcolor{red}{18 months earlier: 76.53\%; 12 months earlier: 79.83\%; } & \\

	\bottomrule
\end{tabular}}
\begin{scriptsize} 
\caption{ \emph{Identified brain dementia classification (diagnosis) and outcome prediction (prognosis) methods published at Medical Image Computing and Computed-Assisted Intervention (MICCAI) from 2010-2016}. We reviewed pioneering works published at MICCAI conference, which typically get extended into journal papers. We report the classification/prediction accuracy for each paper. The majority of these works focused on classification tasks, e.g., classifying Alzheimer's disease (AD) patients vs. normal controls (NC) or stable/progressive mild cognitively impaired (sMCI/pMCI) vs. converted MCI (cMCI) patients. All these works proposed image-based or functional/structural network-based techniques. Notably, very few recent works \cite{Thung:2015,Zhu:2016} presented novel methods for prediction MCI late outcome using magnetic resonance images (MRI) acquired earlier before the outcome measurement. More importantly, none of these works: (1) used morphological networks, which were recently introduced in \cite{Lisowska:2017} for early MCI diagnosis, or (2) proposed a technique for predicting the full trajectory of brain shape changes as MCI progresses towards AD, remains stable, or reverses to normal. This gap needs to be filled, considering there are studies that indicate morphological features of the brain, such as cortical thickness, can be affected in neurological disorders, including AD \cite{Querbes:2009,Brown:2016}. As such, the use of advanced network and shape analysis methods, using machine learning, could prove fruitful for both classification \cite{Lisowska:2017,Mahjoub:2018,Lisowska:2018,Raeper:2018} and prediction tasks. \label{tab:0}} 
\end{scriptsize} 
\end{sidewaystable}

\section{Conclusion}


In this review paper, we examined neuroimaging-based methods for dementia diagnosis and prognosis published in MICCAI 2010-2016 proceedings. The majority of reviewed studies focused on NC, MCI and AD classification tasks using image-based methods or network-based methods including structural and functional brain networks. We noted that very few works developed frameworks to predict MCI conversion to AD at later observations. While the ultimate goal of classification is to provide a computer-aided diagnosis for better clinical decisions, predicting future progression of early demented brains from a baseline observation (i.e., a single timepoint) remains a priority as it might help delay conversion from MCI to AD when early treatment is addressed to the patient. Undoubtedly, predictive intelligence for early dementia diagnosis is still lagging behind, holding various untapped potentials for translational medicine. 

\bibliography{Biblio4}
\bibliographystyle{splncs}
\end{document}